\documentclass[aps,pre,floats,twocolumn,twoside,floatfix,superscriptaddress]{revtex4}

\usepackage{graphicx}
\usepackage{amsmath,amssymb}
\usepackage{psfrag}
\usepackage[utf8]{inputenc} 

\usepackage{color}
\usepackage{hyperref}
\usepackage[usenames,dvipsnames,table]{xcolor}

\newcommand{\Tc}{T_{\scriptstyle \rm c}}
\newcommand{\Tf}{T_{\scriptstyle\rm F}}
\newcommand{\tp}{t_{\scriptstyle \rm p}}
\newcommand{\tpum}{t_{\scriptstyle{\rm p}_1}}

\begin{document}
%\title{On the onset of the percolating cluster in the Ising model after a temperature quench}
%\title{{The role of different spin flips} on the onset of the percolating cluster in the Ising model after a temperature quench}
\title{Energy-lowering and constant-energy spin flips: emergence of the percolating cluster in the kinetic Ising model}

\author{Amanda de Azevedo-Lopes}
\email{azevedo@evolbio.mpg.de}
\affiliation{Instituto de F\'\i sica, Universidade Federal do
Rio Grande do Sul, CP 15051, 91501-970, Porto Alegre RS, Brazil}
\affiliation{Department of Evolutionary Theory, Max Planck Institute for Evolutionary Biology, 24306 Plön, Germany}

\author{Renan A. L. Almeida}
\email{renan.lisboaufv@gmail.com}
\affiliation{Instituto de Física, Universidade Federal Fluminense, Av. Litorânea s/n, 24210-340 Boa Viagem, Niterói, RJ, Brazil}
\affiliation{Departamento de Ciências Exatas, Universidade do Estado de Minas Gerais, Santa Emília, 36800-000, Carangola, MG, Brazil}

\author{Paulo Murilo C. de Oliveira}
\email{pmco@if.uff.br}
\affiliation{Instituto de Física, Universidade Federal Fluminense, 
Av. Litorânea s/n, 24210-340 Boa Viagem, Niterói, RJ, Brazil}
\affiliation{Instituto Nacional de Ciência e Tecnologia - Sistemas
Complexos, Rio de Janeiro RJ, 22290-180, Brazil}

\author{Jeferson J. Arenzon}
\email{arenzon@if.ufrgs.br}
\affiliation{Instituto de F\'\i sica, Universidade Federal do
Rio Grande do Sul, CP 15051, 91501-970, Porto Alegre RS, Brazil}
\affiliation{Instituto Nacional de Ciência e Tecnologia - Sistemas
Complexos, Rio de Janeiro RJ, 22290-180, Brazil}

\date{\today}

\begin{abstract}
After a sudden quench from the disordered high-temperature $T_0\to\infty$ phase to a final temperature well below the critical point $\Tf \ll T_c$, {the non-conserved order parameter dynamics} of the two-dimensional ferromagnetic Ising model on a square lattice (2dIM) initially approaches the critical percolation state before entering the coarsening regime. 
This approach involves two timescales associated with the first appearance (at time $\tpum>0$) and stabilization (at time $\tp>\tpum$) of a giant percolation cluster, as previously reported. 
However, the microscopic mechanisms that control such timescales are not yet fully understood.
In this paper, in order to study their role on each time regime after the quench ($\Tf=0$), we distinguish between spin flips that decrease the total energy of the system from those that keep it constant, the latter being parametrized by the probability $p$.  
We show that observables such as the cluster size heterogeneity $H(t,p)$ and the typical domain size $\ell (t,p)$ have no dependence on $p$ in the first time regime up to $\tpum$. { Furthermore, when energy-decreasing flips are forbidden while allowing constant-energy flips, the kinetics is essentially frozen after the quench and there is no percolation event whatsoever. Taken together, these results indicate} that the emergence of the first percolating cluster {at $\tpum$} is completely driven by energy decreasing flips. 
On the other hand, the time for stabilizing a percolating cluster is controlled by the acceptance probability of constant-energy flips: $\tp(p)\sim p^{-1}$ for $p\ll 1$ (at $p=0$, the dynamics gets stuck in a metastable state). 
These flips are also the relevant ones in the later coarsening regime where dynamical scaling takes place.
Because the phenomenology on the approach to the percolation point seems to be shared by many 2d systems with {a non-conserved order parameter dynamics (and certain cases of conserved ones as well)}, our results may suggest a simple and effective way to set, through the dynamics itself, $\tpum$ and $\tp$ in such systems.

\end{abstract}

\maketitle
\section{Introduction}
\label{sec.intro}

After being equilibrated at a high temperature ($T_0\to\infty$) and suddenly quenched to a temperature well below the critical one, $\Tf \ll\Tc$, the non-conserved order parameter dynamics of the ferromagnetic 2d Ising model on a square lattice first approaches the percolation critical point~\cite{ArBrCuSi07,SiArBrCu07,BlCoCuPi14,BlCuPiTa17}, to then follow a curvature-driven, coarsening dynamics in which the spatial structure is statistically invariant (dynamical scaling~\cite{Bray94}) once distances are measured with the lengthscale $\ell (t)$ associated with the clusters linear size. 
These two temporal regimes are connected, what has led to several predictions that were experimentally verified in liquid crystals~\cite{SiArDiBrCuMaAlPi08,AlTa21} (see the latter reference for a recent and more detailed report on the universal features of the curvature-driven dynamics in these experiments).
Moreover, sharing the geometrical properties of the critical percolation state has a fundamental influence on the late stages of the dynamics, where the asymptotic state may be fully magnetized or divided into parallel stripes~\cite{DeOlSt96,Lipowski99,SpKrRe01a,SpKrRe01b,BaKrRe09,OlKrRe12}. 
The late-time growth of order during the coarsening regime is clearly different from that ruling the development of the stable, giant cluster that characterizes the critical percolation threshold. 
{The moment in which a cluster first percolates defines the timescale $\tpum>0$ (notice that for the triangular lattice, not considered here, $\tpum=0$).}
This first percolating cluster, however, is not stable against the fluctuations still present in the dynamics, despite the absence of activated processes.
It will break and regrow multiple times while competing with the second largest cluster, until  at least one of them stabilizes at the timescale $\tp \sim L^{z_p}$~\cite{BlCoCuPi14,BlCuPiTa17} when the system approaches the percolation critical point.
{ The stabilization of the largest cluster at $\tp$ means that this cluster will grow until a fully magnetized state or a striped one with flat boundaries is attained.}
It is only after this time that the asymptotic power-law behavior associated with the curvature-driven growth starts~\cite{BlCuPiTa17}. 
{ The exponent $z_p$ encodes the dependence on static lattice properties (size and geometry) and, for the square lattice, it has been argued that $z_p=2/5$~\cite{BlCuPiTa17}.
Nonetheless, the dynamical microscopic mechanisms are not well known, neither how they set those timescales.
The  main objective of this paper is to disentangle the role of different classes of spin flips,  regarding their energy changes, on setting the characteristic times for building and stabilizing the first percolating cluster after the temperature quench.}

The geometric description of the 2d ferromagnetic Ising model equilibrium states is relatively simple at temperatures well below and well above $\Tc$. 
In these temperature regimes, connected groups of parallel spins (clusters or domains) have a small size diversity (as defined below). 
At temperatures $T\ll\Tc$,  the system is dominated by a single large, percolating cluster. 
Due to the weak thermal fluctuations, there is also a few small clusters, whose size distribution decays exponentially. 
In the other temperature limit, $T\gg \Tc$, large domains are destabilized by the thermal noise and the system is populated by small clusters whose size distribution is, once again, a decreasing exponential. 
It is only close to $\Tc$ that the diversity of sizes increases and the size distribution broadens, approaching a power-law.
In the thermodynamic limit, $L\to\infty$, the domain size distribution for a single sample becomes dense at all temperatures, i.e., the position of the first missing cluster size also diverges. 
In contrast, for a finite sample, the distribution is dense only below a certain cluster size, and sparse above it.
The diversity of the clusters may be quantified by the so-called cluster size heterogeneity $H$, the average number of distinct cluster sizes occurring in a finite sample, irrespective of the number of domains that are equally sized. 
This observable was studied in percolation~\cite{LeKiPa11,NoLePa11,OlStAr22}, spin equilibrium models~\cite{LvYaDe12,JoYiBaKi12,RoOlAr15} and recently extended to out-of-equilibrium~\cite{AzRoOlAr20} and  more general contexts~\cite{MaAzArCo21}.
The rich behavior of the time-dependent heterogeneity, $H(L,t)$, confirms that it may be useful to understand the interplay between percolation and coarsening dynamics. For simplicity, from now on we will omit the size dependence and write $H(t)$ (analogously for its equilibrium counterpart).

Quenching the system from $T_0\to\infty$, the dynamical heterogeneity $H(t)$ presents a very pronounced peak at the beginning of the dynamics~\cite{AzRoOlAr20}. 
After the quench, correlations start building larger clusters, the domain-size
distribution widens and the growth of $H(t)$ is a consequence of the larger size diversity.
At the same time, the first spanning cluster starts building up, taking most of the system size.
Because less space remains for the other clusters, their number and size decrease, and along with them, the system diversity. 
These two competing mechanisms explain the peak of $H(t)$ occurring slightly before $\tpum$, the time when the first percolation event happens~\cite{BlCoCuPi14,BlCuPiTa17}.
Interestingly, an analogous precursor behavior was also observed for the Voter model~\cite{AzRoOlAr20}, whose dynamics, without surface tension, is much slower and the timescales are longer~\cite{BeFrKr96,TaCuPi15}.
In all these cases, the dynamical heterogeneity $H(t)$ presents three well separated regimes: a peak soon after the quench in temperature, followed by an incipient plateau and, eventually, the power-law decaying. 
These regimes correspond, respectively, to the appearance of the first percolating cluster, its stabilization (and the approach to the percolation critical point) and, finally, the coarsening, curvature driven growth~\cite{AzRoOlAr20}. 
The approximated analytical results of Ref.~\cite{AzRoOlAr20} well explained the height of the plateau and the exponent of the later regime. 
They are, however, valid only after the dynamics has approached the percolation critical point and do not explain the peak of $H(t)$ and the region around it. 
Particularly interesting is the universality of its height: not only it is the same for both the Ising and the Voter model after the quench but also very close to the maximum found in Ref.~\cite{RoOlAr15} for the equilibrium heterogeneity at a temperature well inside the paramagnetic phase ({there are two peaks, one small, very close to $T_c$, and a larger one at a temperature $T_2>T_c$}). 
%As will be discussed in the next sections, other instances of this universality appear related with the nature of different spin flips.

We are interested in the role played by the microscopic mechanisms in building and stabilizing the percolating structures in the early time regime. 
If we take the temperature quench to $\Tf=0$, only spin flips that either decrease ($\Delta E< 0$) or keep the energy constant ($\Delta E=0$) are allowed. 
In particular, how does the characteristic time to attain the percolation critical point depends on these processes? 
Since in the curvature-driven regime domain coalescence does not occur, how does the percolating cluster first appear?
In the next section we describe the model and the methods in more detail while in Sec.~\ref{sec.results} we present new results for the size heterogeneity $H(t)$ and the average linear size $\ell (t)$ of the geometric clusters (connected regions of aligned spins).
Finally, in Sec.~\ref{sec.conclusions} we summarize and discuss our results.

\section{Model and methods}
\label{sec.model}

Starting from an initial state with uncorrelated spins (infinite temperature) having an equal probability of being $\pm 1$, we study the ferromagnetic 2d Ising model after a quench to $\Tf=0$. The
Hamiltonian is
\begin{equation}
  {\cal H}= -J\sum_{\langle ij \rangle} \sigma_i\sigma_j,
\end{equation}
where $J>0$, $\sigma_i$ is the spin at site $i$ and the sum is over all nearest neighbors pairs on an $N=L\times L$ square lattice with periodic boundary conditions. 
After the quench, the system evolves through a continuous time, single-spin  dynamics~\cite{NeBa99}.

In the zero-temperature dynamics, a site is randomly selected and the local field (the sum of the spins on the neighboring sites) is computed.
If, by flipping, this spin becomes aligned with the local field, the energy change is $\Delta E<0$ and the movement is accepted with probability 1. If $\Delta E>0$, i.e., the spin becomes anti-aligned with the local field, the movement is rejected and the spin is left unchanged. 
When the flip does not change the energy of the system, $\Delta E=0$, it is accepted with probability $p$ in order to assess the role of each type of flip. 
For the heat bath and Glauber dynamics, $p=1/2$, while for the Metropolis algorithm, $p=1$~\cite{NeBa99}.
In the constrained case $p=0$, only energy-lowering movements are accepted~\cite{LeDe01,PrBr01,GoLu05}. 
In the continuous-time dynamics~\cite{NeBa99}, we keep a list of the $N_\text{mobile}$ spins that may flip, i.e., those with $\Delta E\leq 0$ and, after each attempt of flipping one spin from this list, the simulation time, measured in Monte Carlo steps (MCS), is incremented by $1/N_\text{mobile}$ and the list is updated.
We present the results for $L=1280$ although simulations were performed with linear sizes up to $L = 5120$.
Averages up to 10000 samples were taken for the smaller systems, while larger sizes require fewer samples (1000).

{Once the percolating cluster is formed at $\tpum$ we follow the two largest clusters.
Since they occupy a large fraction of the system, their magnetizations have opposite signs. 
As they both keep swapping places while competing for space, we measure~\footnote{In the Supplemental Material [URL will be inserted by publisher], a time-lapse of the non-conserved order parameter dynamics of the 2d ferromagnetic Ising model on a square lattice after a quench from a disordered high-temperature state $T_0\to\infty$ to $T=0$. Energy-constant flips are accepted with probability $p$. Spins $+1$ and $-1$ are indicated in light blue, light red, respectively. At each time, the largest cluster (LC) and second largest cluster (SLC) are highlighted, with the colors indicating their magnetization (red colors, $m>0$, blue, $m<0$).} the stabilization time $\tp$ as the last change of sign of the largest cluster~\cite{BlCoCuPi14,BlCuPiTa17}. }

\section{Results}  
\label{sec.results}

 %
%%%%%%%%%%%%%%%%%%%%%%%%%%%%%%%%%%%%%%%%%%%%%%%%%%%%%%%%%%%%%%%%%%%%%%%%%%%%%%%%%%%%%%
%
\begin{figure}[!htb]
\includegraphics[width=\columnwidth]{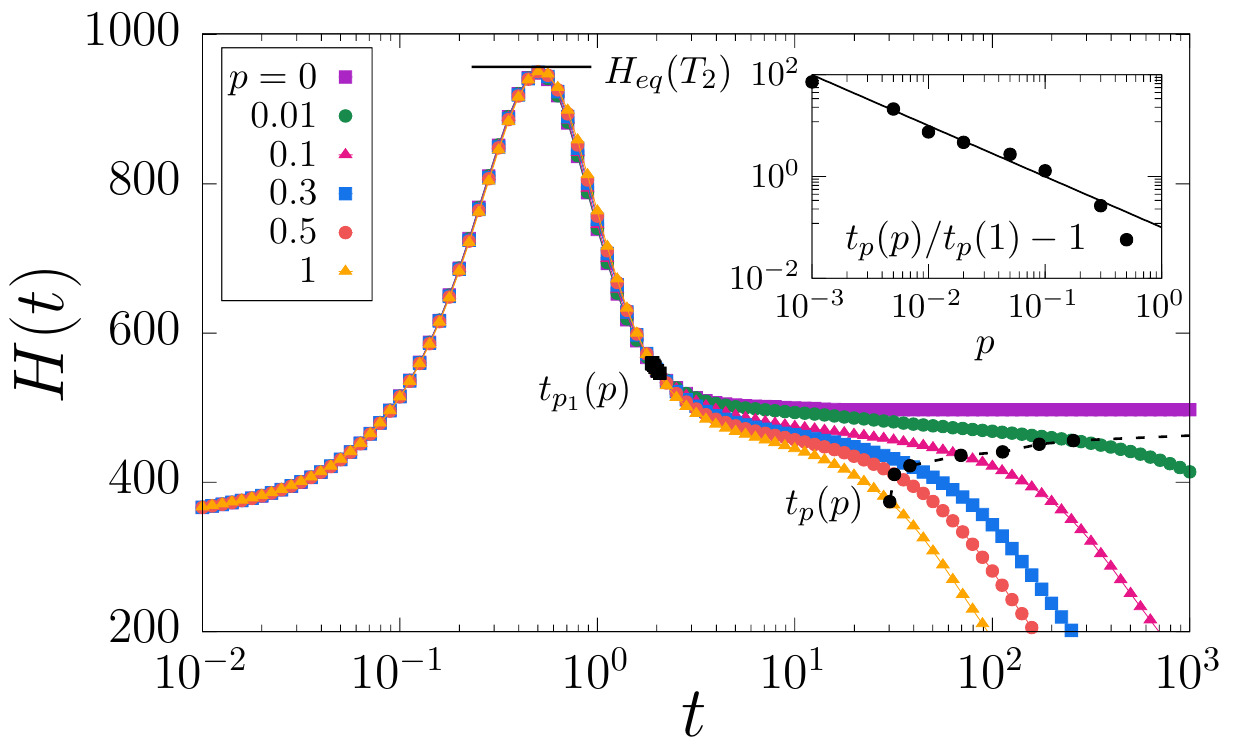}
  \caption{Dynamical cluster heterogeneity, $H(t)$ versus time (in MCS) after a quench from $T_0\rightarrow\infty$ down to $\Tf=0$, for several values of $p$ (the flipping probability when $\Delta E=0$). The system linear size is $L = 1280$. The time a percolating cluster first appears is indicated by $\tpum(p)$ (dark squares), with almost no dependence on $p$. The time that the percolating cluster stabilizes, $\tp(p)$ (dark circles), on the other hand, decreases with $p$. In the first regime, $t\lesssim\tpum$, all curves collapse and $H(t)$ is independent of $p$, differently from the interval $t>\tpum$, where there is a strong dependence on it. 
{ Remarkably, the height of $H(t)$ coincides with the height of the peak for the equilibrium heterogeneity $H_{\rm eq}$ that appears~\cite{RoOlAr15} at a temperature $T_2>T_c$ (indicated by a horizontal line).} The inset shows $\tp(p)/\tp(1)-1$, that seems to have a power-law dependence on $p$ as $p$ decreases,  $p^{-1}$ (solid line). However, above $p\sim 0.5$, $\tp(p)$ barely depends on $p$ and $\tp(p)\simeq\tp(1)$. 
  }

\label{fig.het}
\end{figure}
%
%%%%%%%%%%%%%%%%%%%%%%%%%%%%%%%%%%%%%%%%%%%%%%%%%%%%%%%%%%%%%%%%%%%%%%%%%%%%%%%%%%%%%%
%

The initial, equilibrium state at $T_0\to\infty$ has uncorrelated spins forming small domains that are similar in size. 
The corresponding heterogeneity is $H(t=0) = H_{\rm eq}(\infty)$, where $H_{\rm eq}(T)$ is the equilibrium heterogeneity at the temperature $T$. 
This initial state has a large number of spins with at least half of their neighbors oriented in the opposite direction, i.e., such that putative flips have $\Delta E\leq 0$.
These spins are associated with rough surfaces that may become smoother and with small internal clusters that may disappear (with less holes, the embedding clusters get more compact) or coalesce after a quench. 
With the growth of the domains, their size distribution becomes wider and, consequently, $H(t)$ increases.
Soon after the quench from $T\to\infty$ to zero temperature, there is a peak in $H(t)$~\cite{AzRoOlAr20} as shown in Fig.~\ref{fig.het} for several values of $p$ and $L=1280$. 
The peak occurs within the first MCS, just before a percolating cluster appears for the first time at $t=\tpum$. 
As observed in Ref.~\cite{AzRoOlAr20}, the height of the peak has some degree of universality.
{ Besides being very close to the height of $H(t)$ for the Voter model, it is also close to the largest peak of the equilibrium measure $H_{\rm eq}(T)$ for the Ising model that appear for a temperature $T_2>T_c$.  
Indeed, $H_{\rm eq}(T)$ presents two maxima. 
The first one is located very close to $T_c$ while the second one is larger and is at a temperature well above $\Tc$~\cite{RoOlAr15}, $T_2$. Its height is indicated,} in Fig.~\ref{fig.het}, by a small horizontal line.
In Ref.~\cite{MaAzArCo21} it was argued that an upper bound for $H$ is obtained when $N$ is divided into smaller domains in a deterministic way: one cluster of each size, starting from 1, until the system is fully covered. Obviously, for stochastic systems this bound is never approached either because of repeated sizes or too large clusters that decrease their total number. 
The fact that the heterogeneity, in the several situations described above, reaches a similar maximum is related to getting as close as possible of this upper limit in the presence of stochasticity. 
Also remarkable is that not only the height of the peak, but the whole region up to $\tpum(p)$ does not depend on $p$, i.e., on the dynamics. 
Since $p$ is associated with the $\Delta E=0$ flips, this in an indication that the energy-conserving flips, although present, do not seem to play an important role in the early dynamics. 
{ Notice that if no $\Delta E<0$ flips is allowed whatsoever, the dynamics remains almost totally blocked from the beginning (not shown). 
As discussed below, the important role that $\Delta E=0$ flips have in the stabilization of the percolating cluster and in the scaling regime is only possible if the initial removal of dangling ends ($\Delta E<0$ flips) occurred.}

\begin{figure}[hbtp]
  \includegraphics[width=\columnwidth]{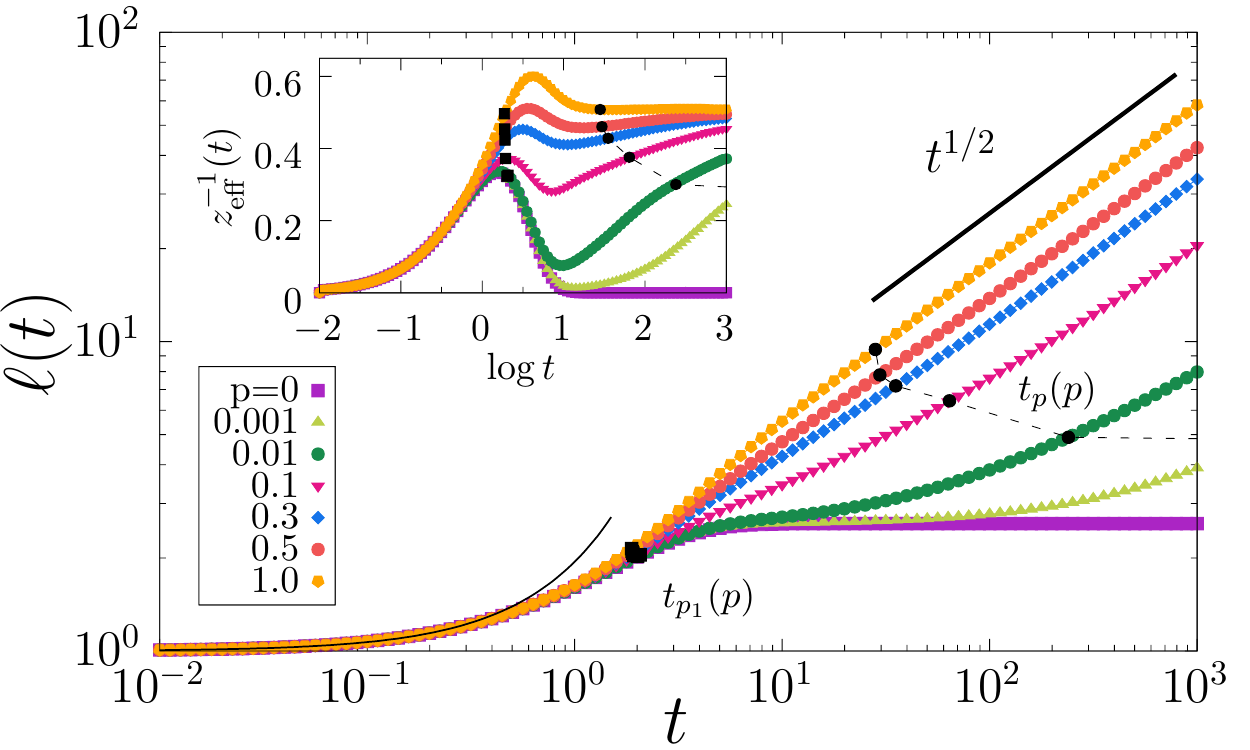}
  \caption{Growing lengthscale $\ell (t)$, as a function of time (in MCS), after a temperature quench from $T_0\rightarrow\infty$ down to $\Tf=0$ for several values of $p$ and $L=1280$. Dark squares and circles locate, respectively, $\tpum(p)$ and $\tp(p)$. The straight solid line indicates the power-law $t^{1/2}$, characteristic of the coarsening regime. The initial exponential growth of $\ell (t)$, indicated by a thin solid line, extends up to the peak of $H(t)$. Inset: local slope, {$z_{\scriptstyle\rm eff}^{-1}(t) \equiv d \log \ell (t)/d \log t$},  as a function of time (semi-log scale).}
\label{fig.Lt}
\end{figure}

%%%%%%%%%%%%%%%%%%%%%%%%%%%%%%%%%%%%%%%%%%%%%%%%%%%%%%%%%%%%%%%%%%%%%%%%%%%%%%%%%%%%%%

For the random initial condition that we consider, the behavior of $H(t)$ strongly depends on $p$ 
between $\tpum(p)$ and $\tp(p)$, the regime in which the percolating cluster stabilizes, and
beyond, Fig.~\ref{fig.het}. 
For $p=0$ the dynamics is halted, never entering the coarsening regime~\cite{OlKrRe11,OlKrRe12}, and $H(t)$ attains a plateau. 
This blocking of the dynamics also occurs in lattices with an odd coordination, as the honeycomb~\cite{TaMi93}. 
Deviations from this plateau are already seen even for very small values of $p$ and the larger $p$ is, the earlier the departure from the plateau occurs.  
Therefore, energy-lowering flips alone are not enough to stabilize the percolating cluster and, eventually, to enter the curvature-driven growth regime. 
As $p$ goes from 0 to 1, increasing the acceptance rate of energy conserving flips, the time interval between $\tp(p)$ and $\tpum(p)$ decreases, i.e., the percolating cluster stabilizes faster.  
The minimum $\tp$ occurs for $p=1$ and the inset on Fig.~\ref{fig.het} shows that although there is little or no dependence for $p\gtrsim 0.5$, below this value it increases very fast, $\tp(p)\sim p^{-1}$.

%%%%%%%%%%%%%%%%%%%%%%%%%%%%%%%%%%%%%%%%%%%%%%%%%%%%%%%%%%%%%%%%%%%%%%%%%%%%%%%%%%%%%%

\begin{figure}[hbtp]
    \centering
    \includegraphics[width=\columnwidth]{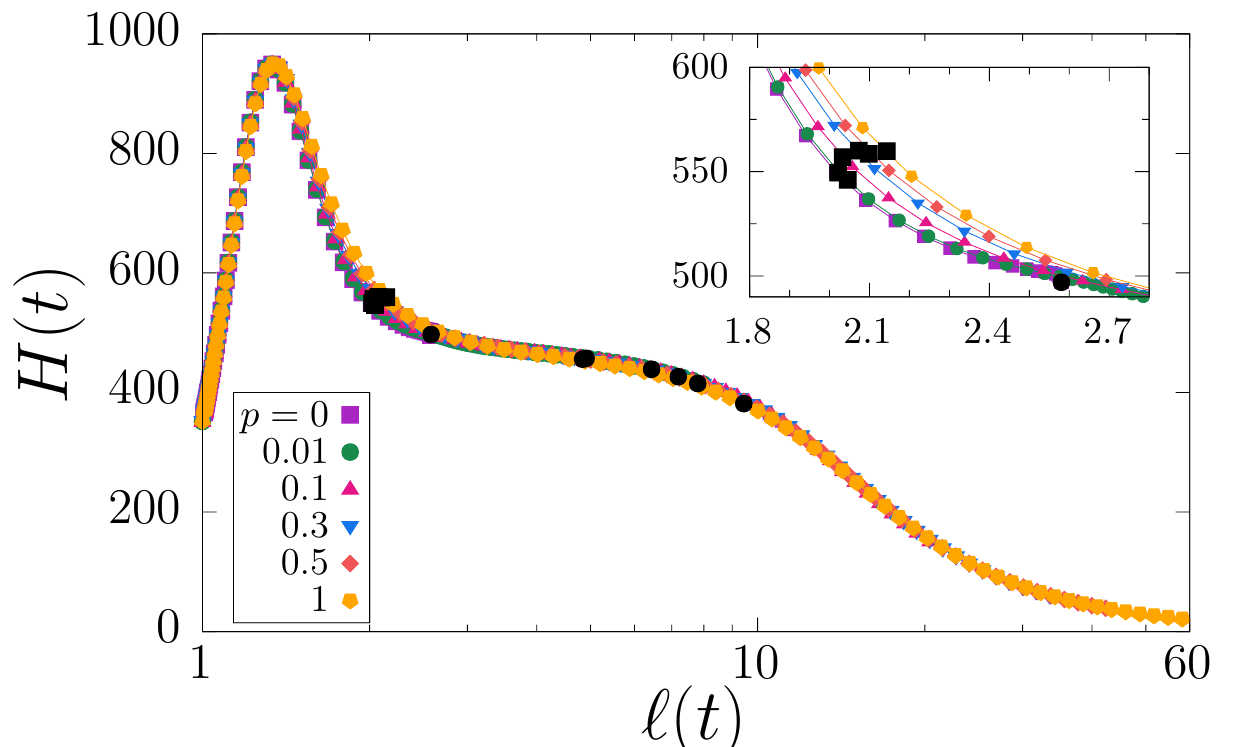}
    \caption{Relation between the $H(t)$ and the growing length $\ell (t)$. Notice that even the $p=0$ case is collapsed onto the universal curve (in this case, however, the curve is interrupted at the maximum value of $\ell (t)$). The scaling is very good, with small deviations occurring close to $\tpum$, as can be seen in the inset.}
    \label{fig:H.L}
\end{figure}

It is also interesting to compare the behavior of $H(t)$ with the growth of a lengthscale associated with the average radius of the domains after the quench, $\ell (t)$.
The main panel of Fig.~\ref{fig.Lt} shows this growing length for several values of $p$ while the inset presents the corresponding instantaneous declivity, i.e., the exponent {$z_{\scriptstyle\rm eff}^{-1}(t) \equiv d \log \ell (t) / d \log t$}. 
Initially, the growth is exponential (indicated by a thin solid line) and, like $H(t)$, $\ell (t)$ has no dependence on $p$ up to $\tpum$ since all curves are collapsed. 
For $p=0$, $\ell (t)$ attains a constant value, whereas the case $p=1$ has the fastest growth.
If $p\neq 0$ and for times larger than $\tp(p)$, the dynamics approach the coarsening regime where $\ell (t)\sim t^{1/2}$. 
The main panel of Fig.~\ref{fig.Lt} shows that, in this late regime, $\ell (t)$ is parallel to the thick solid line while, in the inset, it corresponds to the plateau at $z_{\scriptstyle\rm eff}^{-1}\simeq 1/2$. 
Notice that the coefficient decreases with $p$ and the curves, albeit parallel, are displaced. 
It is in the intermediate regime, roughly between $\tpum(p)$ and $\tp(p)$, that the dependence on $p$ has a larger impact. 
This is an indication of how important the energy-conserving flips are for the stabilization of the percolating cluster. 
For $p\gtrsim 0.5$, the dynamics accelerates after $\tpum$ up to a maximum (e.g., for $p=1$, $z_{\scriptstyle\rm eff}^{-1}\sim 0.6$ at its largest value), and then decreases approaching 0.5. 
Further decreasing $p$, the maximum moves toward $\tpum$ and a minimum develops and becomes deeper, eventually attaining $z_{\scriptstyle\rm eff}^{-1}=0$ as $p\to 0$.

When, instead of the time, we use the characteristic length $\ell (t)$ as the independent variable, Fig.~\ref{fig:H.L}, a very good collapse is obtained for the curves with different values of $p$. Notice that even $p=0$ is included, but in this case the curve is interrupted when the plateau is attained. Some deviations from the scaling are observed around $\tpum$, the beginning of the intermediate region (inset). Before $\tpum$, the collapse is a direct consequence of both $H(t)$ and $\ell (t)$ being independent of $p$, as shown in Fig.~\ref{fig.het}. This is not, however, the case at larger times where there is a strong $p$-dependence and all curves are different. The overall shape of $H(t)$ and the diverse regimes have been discussed in Ref.~\cite{AzRoOlAr20}, including the power-law tail associated with the coarsening regime.

%%%%%%%%%%%%%%%%%%%%%%%%%%%%%%%%%%%%%%%%%%%%%%%%%%%%%%%%%%%%%%%%%%%%%%%%%%%%%%%%%%%%%%

\begin{figure}[htbp]
  \includegraphics[width=8cm,angle=0]{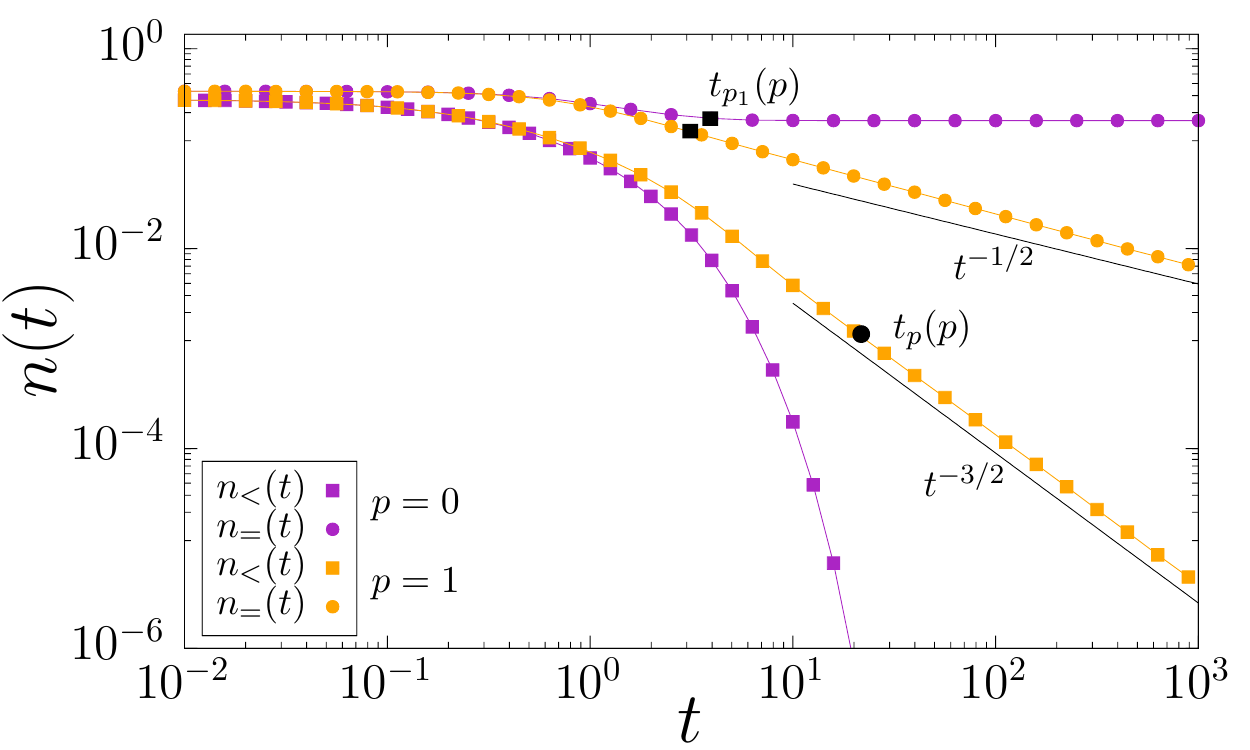}
  \caption{Density of spins whose flip would decrease the energy, $n_<(t)$, or keep it constant, $n_=(t)$, as a function of time, for $p=0$ (dark, purple symbols) and $p=1$ (clear, orange symbols). The system linear size is $L=1280$.  Indeed, coarsening does not occur for $p=0$: $n_<(t)$ vanishes exponentially and the dynamics is blocked soon after $\tpum$. For $p>0$, $n_<(t)\sim t^{-1/2}$ and $n_=(t)\sim t^{-3/2}$ for $t\gg 1$.
 }
    %\textcolor{blue}{Como $n_=(t)\sim t^{-0.53}$ advem das estatisticas de primeira passagem de duas particulas realizando movimento Browniano, seria melhor ja alterar a indicacao para $t^{-1/2}$.}} %% divided by $N=L\times L$ $\neq N_{\Delta E<0} + N_{\Delta E=0}$
    \label{fig.flip}
    \end{figure}

%%%%%%%%%%%%%%%%%%%%%%%%%%%%%%%%%%%%%%%%%%%%%%%%%%%%%%%%%%%%%%%%%%%%%%%%%%%%%%%%%%%%%%

In order to confirm the role of each class of spin flips, we measured the densities $n_<(t)$ and $n_=(t)$ of spins whose flip would either lower or keep the energy constant, respectively. 
As shown in Fig~\ref{fig.flip}, these densities are little affected by $p$ in the beginning of the dynamics, but start to evidence such dependence close to $\tpum$ (for clarity, we only show the extreme cases $p=0$ and 1 where energy-conserving flips are either all forbidden or accepted, respectively). 
For $p = 0$ the dynamics becomes blocked exponentially fast, $n_<(t) \sim \exp(-t/2)$. In contrast, for $p=1$ the asymptotic behavior is a power-law, $n_<(t) \sim t^{-3/2}$ (for intermediate values of $p$, the long-time behavior seems to have the same exponent). 
On the other hand, the density of energy-conserving flips, $n_=(t)$ remains constant, for $t\lesssim\tpum$ even when $p>0$, not because such flips do not occur, but because they correspond only to a shift of the domain walls.
In the long-time limit, $n_=(t)$ attains a plateau for the blocked dynamics ($p = 0$), while $n_=(t) \sim t^{-1/2}$ for $p = 1$. 
The exponent $1/2$ may be understood as follows.
On a square lattice, a circular domain of radius $R$ may be considered as a pile of $2R$ lines whose spins are parallel. 
Each line in the top half is smaller than the one below (in the bottom half is the contrary) and, if it has more than a single spin, only the far-left and/or the far-right spin may flip, with $\Delta E=0$, while all the intermediate spins are blocked with $\Delta E>0$. 
In this way, the overall contribution to $n_=(t)$ is, roughly, $4R$. 
Since the number $N_c$ of clusters linearly decreases with time {(because $N_c\sim A^{-1}$)} while $R$ increases as $t^{1/2}$, one may estimate $n_=(t)\sim R N_c \sim t^{-1/2}$.  
On the other hand, the main contributions to $n_<$ come from large clusters, i.e., those whose areas are much larger than the typical ones ($A\gg\lambda t$, with $\lambda\simeq 2$).
Since the area is much larger than the instantaneous correlation length, many spins in the domain are uncorrelated.
Because of that, these large domains preserve some memory of the random initial state, being less compact and with a rough interface. 
We estimate $n_<(t)$ considering the contributions from dangling ends,
\begin{equation}
    n_<(t) = \int_{A_0}^{L^2} f(A,t) n(A,t) dA,
    \label{eq.n}
\end{equation}
where $A_0$ is a microscopic area and $n(A,t)dA$ is~\cite{SiArBrCu07} the average density of clusters, in the coarsening regime ($t\gg\tp$), with area between $A$ and $A+dA$,
\begin{equation}
    n(A,t)\simeq \frac{2c_d (\lambda t)^{\tau-2}}{(A+\lambda t)^{\tau}},\;\;\; A_0\ll A\ll L^2,
\label{eq.nt}
\end{equation}
$c_d$ is a small constant and $\tau=187/91$. The fraction $f(A,t)$ is the ratio between the average number of spins with $\Delta E<0$ in a cluster of area $A$ at the time $t$ after the quench ($N_<$) and the number of spins in the cluster ($A/A_0$),  $f(A,t)=N_<A_0/A$. The more irregular the perimeter is, the larger $N_<$ is. Area and perimeter are related by $A\sim p^{\alpha}$, where the exponent $\alpha$ is 2 for regular domains while irregular ones have $\alpha<2$.
From this relation, we assume that  $N_<$ is given by deviations from the regularity, i.e., $N_< \sim (p/\ell_0)^{2-\alpha}$, where $\ell_0$ is a microscopic length. Thus
\begin{equation}
    f(A,t) \sim \left(\frac{p}{\ell_0}\right)^{2-\alpha}\frac{A_0}{A}
    \sim A^{1/\alpha}(\lambda t)^{1/2-\alpha},
\label{eq.f}
\end{equation}
where the area-perimeter relation was extended~\cite{SiArBrCu07} to include time,
$$
\frac{A}{\lambda t}\sim \left( \frac{p}{\sqrt{\lambda t}}\right)^{\alpha}.
$$
For large clusters ($A\gg\lambda t$), those that indeed contribute most for $N_<$, $\alpha\simeq 1$~\cite{SiArBrCu07}. Using this value, and substituting Eqs.~(\ref{eq.nt}) and (\ref{eq.f}) into Eq.~(\ref{eq.n}), we obtain $n_<(t)\sim t^{-3/2}$. 
Notice also that when the power-law regime settles ($t>\tp$), $n_<(t)\ll n_=(t)$, the flips that decrease the energy are much less frequent, and thus less relevant in this regime, than those that keep it constant.

Finally, we also studied the different asymptotic states, either the ground state or a set of parallel stripes, and how the corresponding probabilities depend on $p$ (except, of course, for $p=0$).
Within the error bars, our results (not shown) are essentially constant and compatible with those for $p=0.5$~\cite{SpKrRe01a,SpKrRe01b} and
$1$~\cite{BaKrRe09}. % and 0.5~\cite{SpKrRe01a,SpKrRe01b}.

\section{Conclusions}
\label{sec.conclusions}

Soon after the ferromagnetic 2d Ising model is taken out of equilibrium by a quench from $T_0\to\infty$ to $\Tf\to 0$, there appear two timescales, $\tpum$ and $\tp$, respectively associated with the first formation and the subsequent stabilization of the percolating cluster. 
{Although in Refs.~\cite{BlCoCuPi14,BlCuPiTa17} it was shown that both $\tpum$ and $\tp$ depend on the lattice properties, it remained unclear which were the corresponding dynamical microscopic mechanisms.
We here presented an attempt to elucidate these issues by disentangling the spin flips that decrease the energy (always accepted) from those that keep it constant (entropic flips, accepted with probability $p$) under the non-conserved order parameter dynamics.  
There are three relevant time intervals: 
from the quench until the first percolating cluster appears at $\tpum$ ($0<t<\tpum$),
from $\tpum$ until the stabilization of the percolating cluster at $\tp$ ($\tpum<t<\tp$)
and, finally, the scaling regime that may arise beyond $\tp$.}

The short-time dynamics, up to the time $\tpum>0$ when the first percolating cluster appears, is independent of $p$, for the observables that we considered here, the heterogeneity $H(t)$ and the lengthscale $\ell(t)$.
The results are all equivalent, collapsing even when completely suppressing the constant-energy flips ($p=0$). 
Thus, in this initial regime, spin flips with $\Delta E=0$ play no relevant role on the construction of the large, percolating cluster, differently from the coalescence processes driven by the energy-lowering spin flips. 
It is indeed the coalescence of small domains that explain the exponential growth that leads to the first cluster spanning the whole system.
{ Moreover, if the $\Delta E<0$ flips were forbidden while allowing those with $\Delta E=0$, the dynamics would be essentially frozen since the initial disordered state and there would be no percolation event at $\tpum$ whatsoever. 
The energetic flips are thus not only important for the formation of the percolating cluster but also essential for unblocking the $\Delta E=0$ flips that are relevant in the other time intervals.}

{After being formed, the percolating cluster is broken and rebuilt many times until} it becomes stable at a much larger time ($\tp$), when the system approaches the percolation critical point. The larger the probability $p$ of accepting energy conserving flips is, the sooner this stabilization is attained. 
When $p\ll 1$, we obtain $\tp(p)\sim p^{-1}$. 
%Lowering $p$, the dynamics gets slower because fewer proposed zero-energy flips are accepted. 
Without these flips ($p=0$), the system freezes soon after $\tpum$.
This reveals the pivotal role played by the spin flips with $\Delta E = 0$ for the stabilization of the percolating cluster.
{ If the processes leading to $\tp$ occur after the number of energy-decreasing flips became sufficiently small and most of the spins that may flip would keep the energy constant, the probability $p$ is only a rescaling time factor, explaining the $p^{-1}$ behavior. 
In this regime, the curves can be indeed collapsed once time is rescaled by $p$ (not shown).}
Note that $\tp$ depends on the system size as $\tp \sim  L^{z_{\rm p}}$, and it was conjectured that $z_{\rm p} = 2/5$ for the square lattice~\cite{BlCoCuPi14,BlCuPiTa17}. 
It is still an open question whether $z_{\rm p}$ depends on $p$. 
Combining with our result, we may write $\tp(L,p) \sim L^{z_{\rm p}}/p$ { for small $p$}. 

{Once the dynamics enter the curvature-driven regime, the spin flips with $\Delta E=0$ continue to have a central role}.
Indeed, $n_<(t)  <  n_=(t)$ soon after $t_p$ and, as shown by both simulations and theoretical arguments, they asymptotically evolve as $n_<(t)\sim t^{-3/2}$ and $n_=(t)\sim t^{-1/2}$.
%This result, that the timescale associated with the stabilization of the percolating cluster is the inverse of the probability of the constant energy flips occurring, is the confirmation that these flips are the solely responsible for the coarsening process.
{ But despite the fact that the energy decreasing flips are no longer the most relevant in this later regime, their role at the beginning of the dynamics, while the percolating cluster was being formed, is essential for the very existence of the subsequent regimes.
Without the formation of the percolating cluster, the dynamics becomes blocked by the excess of dangling ends that are no longer removed (the process involves energy-decreasing flips). }

%{  In the configuration space, constant-energy paths, responsible for the coarsening behavior, are not allowed unless the energy-decreasing paths had been followed first.}

The importance of the early approach to critical percolation to the ensuing dynamics and the geometric properties of clusters and interfaces have been emphasized and recognized in the last years.
Although we only considered here the non-conserved order parameter dynamics (Model A), the approach to the percolation point also occurs for symmetric mixtures (50:50)  the model B dynamics that conserves the order parameter~\cite{SiSaArBrCu09,TaCuPi16,TaCuPi18}. 
Understanding how $\tp$ depends on $p$ in this case may be relevant for a multitude of phase separating systems. 
Moreover, it is possible to devise protocols that change the value of $p$ along the dynamics. For example, the system may first evolve with a constrained dynamics ($p=0$), commonly observed in glassy systems and colloidal suspensions~\cite{Zaccarelli07}, until a metastable state is attained. From that state on, it may evolve following the chosen $p(t)$ protocol in order to study, for example, the yield to escape from metastable states and the role of memory in the reorganization of the geometrical structures. 
Whatever the dynamics is, it would be interesting to find ways to mimic the effect of $p$ on experimental setups.
In addition, it would be important to access the universality of those results on lattices with different coordination numbers such as the triangular one~\cite{SaSaGr83,ChCuPi21} and on different systems such as the Potts~\cite{ViGu86,DeOlSt96,LoArCuSi10,LoArCu12,DeRe19} and Voter models~\cite{BeFrKr96,TaCuPi15,TaCuPi18,AzRoOlAr20,LaDaAr22}. 

\begin{acknowledgments}
Work partially supported by the Brazilian funding agencies CNPq, CAPES and FAPERJ. JJA thanks the Brazilian funding agency CNPq (Grant No. 308927/2017-6). AAL thanks the Brazilian funding agency CAPES in the framework of the Capes-PrInt program (Grant No. 88887.466912/2019-00).
\end{acknowledgments}

\bibliographystyle{apsrev}   
%\bibliography{het} 

\end{document}